\documentclass[12pt,preprint]{aastex}

\lefthead{Escala}
\righthead{On the global triggering mechanism of star formation in galaxies.}

\def\msun{M_{\odot}}

\begin{document}

\title{On the global triggering mechanism of star formation in galaxies.}

\author{Andr\'es Escala}
\affil{Departamento de Astronom\'{\i}a, Universidad de Chile, Casilla 36-D, Santiago, Chile.}
\affil{Kavli Institute for Particle Astrophysics and Cosmology, Stanford University Physics Department / SLAC, 2575 Sand Hill Rd. MS 29, Menlo Park, CA 94025, USA.}

\begin{abstract}
We study the large-scale triggering of star formation in galaxies. We
 find that the largest mass-scale not stabilized by rotation, a well
 defined quantity in a rotating system and with clear
 dynamical meaning, strongly correlates with the star formation rate
 in a wide range of galaxies. We find that this relation can be
 explained in terms of the threshold for stability and the amount
 of turbulence allowed to sustain the system in equilibrium. Using this relation we also derived  the observed
 correlation  between the star formation rate   and the  luminosity of the
brightest young stellar cluster

\end{abstract}


\section{Introduction}

Observations  of normal spiral galaxies by Schmidt (1959)   first
suggested  that their star formation rates (SFR) scales with the global properties. This conclusion was extended to other
 galaxies with higher SFR, such as the nuclear 
regions of spiral galaxies  and Ultra Luminous InfraRed Galaxies
(ULIRGs) by Kennicutt (1998). These galactic scale observations have
lead to an empirical law for star formation called Kennicutt-Schmidt
(KS) Law:
\begin{equation}
 \rm \dot{\Sigma}_{star} \propto \Sigma_{gas}^{1.4} \,\,\, ,
\end{equation} 
where $\rm  \Sigma_{gas}$ and $\rm \dot{\Sigma}_{star}$ are the gas
surface density and SFR per unit area. 

Since star formation is a local
process that happens on subparsec scale, the correlation with global
(galactic scale) quantities such as averaged $\rm  \Sigma_{gas}$,  suggest the
 existence of a physical connection between galactic ($>$ 1 kpc) and
subparsec scales. Motivated by the existence of the  KS law,  several 
authors have tried to find the link in which a global/large scale property of  
galaxies could trigger and/or regulate  star formation in them  (Quirk
1972; Wyse 1986; Larson 1987; Kennicutt 1989; Silk 1997; Tan 2000; 
Elmegreen 2002; Li et al. 2005). It is also important to note that the  KS 
law can still  be explained in terms of mainly local processes within 
starforming clouds (Krumholtz et al. 2009 and references therein).

 From the study of   gravitational
instabilities in disks, the  Toomre  parameter $\rm Q \equiv C_{\rm s}
\kappa / \pi G \Sigma$ (or
 variations of that such as the star formation threshold $\rm
 \Sigma_{crit}$; Kennicutt 1989)
 arises as the most natural candidate as key triggering
 parameter. However, the average Q in a galaxy is 
observed to be  never too far from 1,  from local spiral galaxies (Martin \& Kennicutt 2001), to starbursts such as ULIRGs (Downes \& Solomon 1998). This is 
believed to be due to self-regulating feedback processes in the
following way (Goldreich \& Lynden Bell 1965):
if Q $>>$ 1, then the disk will cool rapidly and form stars, while if
Q $<<$ 1, then the star formation will be so efficient that the disk
will heat up to Q $\sim$ 1. Observed variations of Q (or $\rm
 \Sigma_{gas}/\Sigma_{crit}$) are  at best within 
factors of a few, which makes  hard to explain the variations of 6-7
orders of magnitude that varies  the SFR in galaxies, solely in terms of
this threshold. 

However, for different disks all at the condition of   marginal Toomre
stability,  is still possible to have  different equilibrium states and
this is the topic that  we will explore in this $Letter$. The aim of
this work, is to study  why the SFR in
some galaxies can be orders of magnitude higher than in other ones,
more specifically, which galactic property triggers this behavior.

This work is organized as follows. Starts with 
a review  the  gravitational instability analysis with a focus on the
role of turbulence in \S 2. In \S 3, continues studying the effects of
the global threshold for stability on the star formation
activity, and the implications of this. Finally in \S 4, we discuss the  results of this $Letter$.

\section{GRAVITATIONAL INSTABILITY IN REAL ISM AND THE ROLE OF TURBULENCE}

We start reviewing  some standard results from gravitational
instability analysis (Toomre 1964; Goldreich \& Lynden Bell 1965). For
one
of  the simplest cases of a differentially
rotating thin sheet or disk, linear stability analysis of such a
system yields the  dispersion relation for small perturbations (Binney
\& Tremaine 2008) of $\omega^2 = \kappa^2 - 2\pi G \Sigma |k| + k^2 C_{\rm s}^{2} $. Where $C_{\rm  s} = \sqrt\frac{dP}{d\Sigma}$ is the sound speed,
$\Sigma$ is the surface density, and $\kappa$ is the epicyclic frequency
, which is given by $\kappa^2 (R) = R \frac{d\Omega^2}{dR} \,
+ \, 4 \Omega^2$ (being $\Omega$ the angular frequency). The system becomes unstable when $\omega^2 < 0$ which is equivalent
to the condition   Q $<$ 1, where Q is the Toomre parameter and is 
defined as  $\rm  C_{\rm s} \kappa / \pi G \Sigma$ . In such a case, there
is a range of unstable length scales limited on small scales by
thermal pressure (at the Jeans length $\lambda_{\rm Jeans} = C_{\rm s}^2 / G\Sigma $) and on large scales by rotation (at the critical length set by rotation, $\lambda_{\rm rot}= 4\pi^2 G \Sigma / \kappa^2$).  All intermediate length scales are unstable, and the most rapidly growing mode has a wavelength $2 \, \lambda_{\rm Jeans}$.

The maximum unstable length scale in a disk, $\lambda_{\rm rot}$, is a robust quantity because it depends only on the surface
density and epicyclic frequency of the disk and not on the smaller scale
physics. Such scale has a characteristic mass associated, defined
equals to  $ \Sigma_{gas} (\lambda_{\rm rot}/2)^{2}$ and that can be expressed as:
\begin{equation}
\rm M_{\rm rot} = \frac{4\pi^4G^2\Sigma_{\rm gas}^3}{\kappa^4} \, .
\label{mrot}
\end{equation}

Contrarily, due to the  complex structure and dynamics
of real interstellar medium (ISM) in galaxies, which cannot be described by a simple equation of state,
there is not a
well-defined  Jeans length at intermediate scales. Therefore, no real lower
limit on the sizes of the self-gravitating structures that can form
until the  thermal Jeans scale is reached in molecular cloud cores
(Escala \& Larson 2008).

One of the intermediate scales  usually claimed to be a relevant in
the stability of the ISM, is
the Jeans length defined by the turbulent pressure in the medium
($\lambda_{J}^{turb}= v_{\rm turb}^2 / G\Sigma $).  Besides the convenience of the idea of a
turbulent pressure term that generalize  the gravitational instability
analysis, this term is  however  not a well defined quantity in ISM (Elmegreen \& Scalo 2004). This comes  from the  fact 
that  this  pressure term could only  be defined  in the case
where the dominant turbulent scale is much smaller than the region under
consideration (`microturbulence'), which is in fact is not the case of
ISM. Rigorous analysis endeed shows that turbulence can be represented as a pressure only in the
microturbulence case (Bonazzola et al. 1992). Besides turbulent
motions  could help to prevent  collapse on some scales, those
complex motions cannot be represented as a  pressure. Therefore the
gravitational instability  analysis is not strictly applicable with a
turbulent pressure term that  could stabilize and  damp all the
substructure below $\lambda_{J}^{turb}$. 

Although turbulence cannot be modelled as simple pressure supporting
term that will prevent all gravitational instabilities below some
scale, turbulence could prevent the growth of instabilities by heating
the gas trough compression. However,  compression from turbulence not
only heat the gas, but  also may  enhance
 collapse and star formation on smaller scales (Elmegreen 2002; Krumholtz 
\& Mckee 2005).  Only  when the turbulent heating  becomes faster than
the gas cooling, the net
 effect of turbulence is to effectively heat the gas and therefore
 prevent gravitational instabilities. Once this condition is satisfied
 the gas is heated towards  $\rm c_{S} \sim v_{turb}$, the thermal
 pressure will  stabilize  and  damp all the substructure below
 $\lambda_{J} \sim  \lambda_{J}^{turb}$. If the condition  $\lambda_{J} \sim \lambda_{rot}$
 (equivalent to 
 Q $\sim$ 1) is also satisfied, the system will then became globally stable. 

A well studied  case in which both conditions could be satisfied, 
 is  in self-gravitating protoplanetary disks where the heating is due  to  gravito-driven turbulence. Gammie (2001)  showed that turbulence heat the gas such  the
 system becomes stable  on all scales (Q $>$ 1), when $\rm t_{cool} \,
 > \, 3 \, t_{dyn}$, being  $\rm t_{cool} \,\,  and \,\,  t_{dyn}$ the
 cooling and orbital times. Such disk is in an equilibrium state
 (so-called `$gravitoturbulence$') that 
 experiences significant fluctuations, but the disk is stable against
 fragmentation and maintains itself in the brink of instability (Rafikov 2009)

Galaxies are indeed observed to be close to equilibrium (Martin \&
Kennicutt 2001; Downes \& Solomon 1998), with observed
Toomre Q parameters  never too far from 1 (averaged over the
whole system and using the turbulent version of the Toomre parameter; $\rm  Q_{turb} = v_{\rm turb} \kappa / \pi G \Sigma$). This is due to 
self-regulation  heating/cooling processes: if $\rm Q_{turb} >>$ 1, in
the absence of heating driven by  instabilities the disk will cool rapidly and the system will
eventually become unstable, while if  $\rm Q_{turb} <<$ 1, then the self-gravity  and
star formation feedback will be so efficient that will produce enough
turbulence to heat the disk towards  $\rm Q_{turb} \sim$ 1. 

However, galaxies are in an  state that departs somewhat from the
`$gravitoturbulent$' one found by Gammie (2001). Because in galaxies
only the turbulent Q Toomre parameter is close to one, not
the thermal Q which is the one that guarantees stability (like in `$gravitoturbulent$' state). Galaxies are
only close to stability, the runaway growth of density fluctuations is not suppressed and  the formation of
bound objects on different scales is ongoing (i.e. star formation, GMC
formation, etc). They are probably oscillating around marginal
stability due to the self-regulation feedback process, in a more
dynamical fashion (and with larger oscillations) than the one studied in Gammie (2001). 

Besides this  more complex behavior of ISM in galaxies compared with
the classical self-regulated `$gravitoturbulent$' state studied in
protoplanetary disks, the threshold is still well defined and there is
self-regulation processes toward this marginal state. The marginal
 stability is well defined at thermal Q=1 and this happens when the
 following  two conditions are satisfied: $\rm Q_{turb} \sim 1$ and 
$\rm t_{cool} > t_{heat}$.

\section{Global Threshold for Stability and its Effect on the Star Formation Activity}

Besides most disks in galaxies  are in the same state of being close to marginal Toomre stability, still
 some disks like nuclear disks in starbursts are able to be much 
 more turbulent than others such as the disks of spiral galaxies. The
 reason  is that  is still possible to have  different equilibrium solutions 
for $\rm Q \sim 1$, which is equivalent to $\rm M_{\rm rot} \sim
M_{\rm Jeans}$. Simply because some disks starts with a higher
threshold for stability (a larger mass-scale not stabilized by
rotation, $\rm M_{\rm rot}$), their large-scale conditions  requires to  the
self-regulation processes to produce more turbulence  in order to heat
 the gas towards this  higher value for  globally stabilize the
 system. Under this view, the self-regulation  drives the  system
 towards an equilibrium with a higher level of turbulence, and this
 more turbulent  state is triggered by the initial condition of
 having a system with  higher threshold for stability.

The existence of different equilibriums in disks is particularly
relevant because is believed that turbulence has a role in
enhancing and possibly controlling star formation (Elmegreen
2002; Krumholz \& McKee 2005; Wada \& Norman 2007). On its simplest 
form (proposed by
Elmegreen 2002) the SFR depends on the probability distribution
function (PDF) of the gas density produced by galactic turbulence,
which appears to be lognormal in simulations of turbulent molecular
clouds and interstellar medium (Wada \& Norman 2001;
Ballesteros-Paredes \& Mac Low 2002; Padoan \& Nordlund 2002; Li et
al. 2003; Kravtsov 2003; Mac Low et al. 2005; Wada \& Norman 2007;
Wang \& Abel 2009). The
dispersion of the lognormal PDF, is believed to be determined by the
rms Mach number  of the turbulent motions  if high-density regions are
formed mainly through shock compression in a system (V\'azquez-Semadeni 1994;
Padoan et al. 1997; Nordlund \& Padoan 1998; Scalo et
al. 1998). Therefore it is expected that the SFR in galaxies 
scales with the velocity dispersion of turbulent motions.

In summary, the existence of a higher threshold  for stability
($\rm M_{rot}$) and the self-regulation towards  marginal stability, allows
the disk to be  in equilibrium with a higher turbulence level, which
itself enhances a higher star formation activity. Therefore, both
galactic-scale processes ($\rm M_{rot}$) and local processes
(turbulence within molecular clouds) are relevant in regulating star
formation. This global well-defined scale in disks, $\rm M_{rot}$,
plays the role of  triggering turbulence and therefore 
the  star formation activity in galaxies.

In order to test if this scenario is correct, we will check  if   the mass-scale
defined by rotation (Eq. 1) correlates with the star formation rate as
expected. For a rotationally supported system, the average of this mass-scale  can be expressed
in terms of quantities such as the total gas mass and gas fraction,
that are easier to compare with observations (Escala \& Larson 2008):
\begin{equation}
\rm M_{rot}  = 3 \,\, 10^7\,\msun \,\, \frac{ M_{\rm gas}}{10^9 \,\msun} \left(\frac{\eta}{0.2}\right)^2,
\label{eq2}
\end{equation}
where $\rm M_{\rm gas}$ is the total gas mass in the disk and  $\rm
\eta = M_{\rm gas}/M_{\rm dyn}$ is the ratio of the gas mass to the
 total enclosed dynamical mass within the gas radius (this varies from
 the disk radius for spiral galaxies, to the radius of the nuclear starburst 
 disk/ring in ULIRGs).

\begin{figure}
\plotone{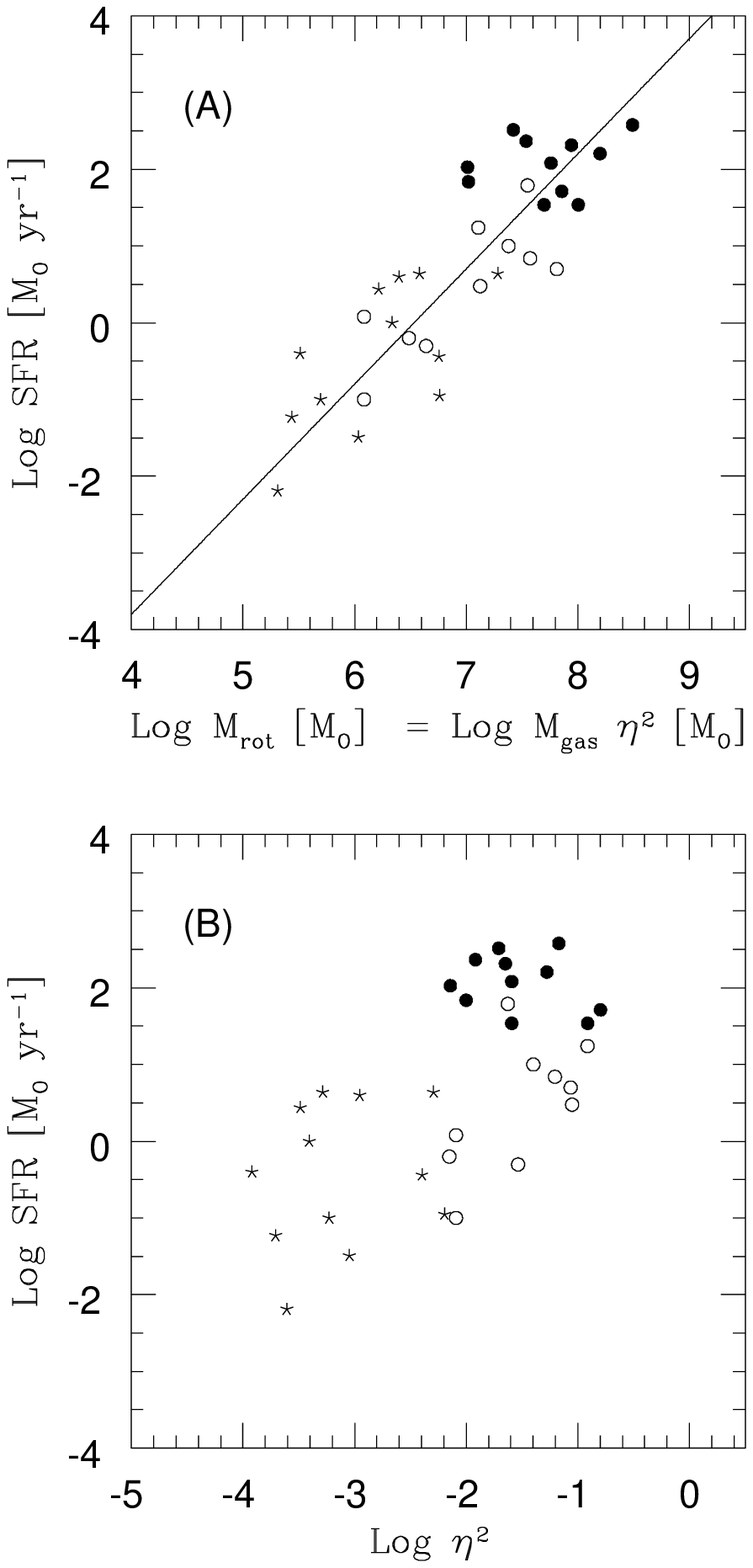}
\caption{a)  The  star formation
 rate  plotted against the  critical mass-scale defined by rotation
 $\rm M_{rot}$,   estimated from Eq. \ref{eq2}
  using measured quantities in those galaxies. The black dots shows data 
for nuclear starburst  disks,
 the stars for normal spirals galaxies and the open circles for
 nuclear gas in spirals. The solid line corresponds to $\rm  SFR \propto M_{rot}^{1.4}$.
b) The  star formation
 rate  plotted against  the square of the gas fraction $\rm
\eta = M_{gas}/M_{dyn}$, using the same symbol nomenclature  as in a).
}
\label{SFR}
\end{figure}

In figure \ref{SFR}a we plot   the stability threshold defined by
rotation estimated from Eq. \ref{eq2}, against the measured star formation rate
in those galaxies. With stars we plot the data taken for normal
spirals (Kennicutt 1998; Kent 1987; Pisano et al. 1998; Giraud 1998; Theis
2001; Braine 2001; Leitherer 2002;
Helfer et al. 2003; Gallagher 2005; Afanasiev 2005; 
Davidge 2006; P\'{e}rez-Torres \& Alberdi 2007; Thilker et al. 2007), with open circles for the nuclear gas in normal
spirals (Jogee 2005; Mauersberger et al 1996; Alonso-Herrero 2001; Hsieh et al 2008) and with full circles for ULIRGs (Downes \&
Solomon 1998). Figure \ref{SFR}a shows a clear correlation between
$\rm M_{rot}$ and the star formation rate, in agreement with the
scenario outlined in this $Letter$. Figure \ref{SFR}a supports   the critical  role of this threshold 
mass in the triggering of star formation. The solid line in figure
\ref{SFR}a represents a
star formation law of $\rm  SFR \propto M_{rot}^{1.4}$.

For  comparison purposes we
also plot in figure \ref{SFR}b the  measured star formation rate
against the square of the gas fraction $\rm
\eta$. Figure \ref{SFR}b clearly shows that the scatter in this
relation is considerably increased compared to  figure
\ref{SFR}a. This means that the gas fraction is not a more
fundamental parameter in controlling the star formation rate. We
do not plot the SFR against the gas mass, since it is well established that
there is no correlation of the total gaseous mass in galaxies with their
 current star formation rate. 



In summary, the predicted correlation between the SFR and the maximum
unstable mass defined by rotation is indeed observed in galaxies, in a
range that spans for 5 orders of magnitude in SFR. This is an strong
suggestion that the global threshold for instability indeed triggers
star formation, by allowing the disk to be in equilibrium configuration with
a higher turbulence level.

\subsection{Relation between the SFR and most luminous young stellar cluster}

Probably the most straightforward application of the scenario outlined in \S 2,
is to check if its able to explain the observed correlation  between
the star formation rate of a galaxy  and the  luminosity of its
brightest young stellar cluster (Larsen 2002; Bastian 2008). This is
because the  mass-scale studied here and that  correlates with the SFR (Fig
1), also corresponds to the
most massive unstable cloud in a disk, which
could lead to the formation of the most massive and luminous young cluster in such
system (Escala \& Larson 2008). 

The total luminosity of a cluster, can be computed for a given initial
mass function (IMF) and mass-luminosity relation. Assuming a Salpeter
IMF $\rm (\frac{dN}{dm} \propto m^{-2.35})$ and the usual mass-luminosity
relation for main sequence stars $\rm (L \propto m^{3.5})$, the total
luminosity of a cluster is given by
\begin{equation}
\rm L_{tot} \propto M_{cloud}^{0.92} \, ,
\end{equation}
where $\rm M_{cloud}$ is the total mass of the parent cloud that
formed the cluster, which  is assumed to satisfy the observed
relation with the most massive star of such cluster: $\rm
M_{star}^{max} \sim M_{cloud}^{0.43}$ (Larson 1982). 

Taking into account the correlation found between the most massive
unstable 
cloud and the star formation rate in galaxies (Fig 1), which is
approximately $\rm M^{max}_{cloud} \propto SFR^{\small \frac{1}{1.4}}$, the total luminosity
of the brightest cluster is given by $\rm L_{tot}^{brightest} \propto
SFR^{0.66}$. Finally, this can be expressed in terms of absolute
magnitude in the  V-band by $\rm M_{V}^{brightest} = 4.79 - 2.5 \, Log
L_{V}^{brightest}$, arising to the relation:
\begin{equation}
\rm  M_{V}^{brightest} \propto -1.65 \,\,  Log \, SFR \, ,
\end{equation}
which is in good agreement with the observed slope of -1.87 (Weidner
et al. 2004). Particularly good if we take into account  all the uncertainties
in the assumptions, such as the slopes of the IMF and  the mass-luminosity
relation.
 




\section{SUMMARY}

In this Letter we have studied gravitational instabilities in disks, with special attention on the
role of turbulence in stabilizing the system. We discussed that although turbulent motions cannot be
modelled as a pressure supporting term, turbulence can still make a disk globally stable when both the
turbulent Toomre (Q$\rm_{turb}>$1) and Gammie ($\rm t_{cool} > t_{heat}$) conditions are satisfied. This allows to the
global threshold for stability, which is defined by largest scale in galactic disks  not stabilized by
rotation, to have a clear role in the dynamics of the ISM.

For a disk with a larger scale not stabilized by rotation, its dynamical equilibrium configuration
(Q$\sim$1) is with a more turbulent ISM.
Therefore, the role of the global threshold for stability in a disk
is to define up to which amount the generation of turbulence is allowed. Since turbulence enhance collapse on small scales and therefore star
formation, on the scenario proposed in this Letter, we expect a
correlation between the mass-scale for global stability and the star formation rate.

We found that this relation is indeed observed in galaxies that ranges from ULIRGs to normal spirals.
Compared with other relations such us Kennicutt-Schmidt Law, its relevance rely on  that is the only
correlation of the  star formation rate with a quantity with clear dynamical
 meaning on galactic
scales.

We also explore the implications of this predicted relation between global stability and the star
formation rate. We found that based on this relation, we are able to explain the observed correlation
between the star formation rate in galaxies and the most luminous young stellar in them.

I would like to thank Richard Larson for valuable comments.  I
acknowledges partial support from Center of Excellence in Astrophysics
and Associated Technologies (PFB 06), from Centro de Astrof\'\i sica
FONDAP 15010003 and from GEMINI-CONICYT FUND.








\begin{thebibliography}{}

\bibitem[Binney \& Tremaine (1987)]{az05} Afanasiev V. L., Sil\'chenko
  O. K., 2005, A\&A,  429,  825 
\bibitem[Binney \& Tremaine (1987)]{az92} Alonso-Herrero A., Engelbracht C. W., Rieke M. J., Rieke G. H., Quillen A. 
C., 2001, \apj,  546,  952
\bibitem[Binney \& Tremaine (1987)]{az98} Ballesteros-Paredes, J., \& Mac Low, M.-M. 2002, ApJ, 570, 734
\bibitem[Binney \& Tremaine (1987)]{bk88} Bastian N., 2008, MNRAS, 390, 759
\bibitem[Binney \& Tremaine (1987)]{bt87} Binney, J., Tremaine,
  S. 1987, Galactic Dynamics. Princeton University Press, Princeton

\bibitem[Binney \& Tremaine (1987)]{ba88} Bonazzola S., Perault M., Puget J. L., Heyvaerts J., Falgarone E., Panis J. 
F.,  1992, Journal of Fluid Mechanics,  245,  1
\bibitem[Downes \& Solomon (1998)]{ds98} Downes, D., Solomon, P. M.  1998, \apj, 507, 615 
\bibitem[Binney \& Tremaine (1994)]{dr99} Davidge T. J., 2006, \apj,  641,  822
\bibitem[Binney \& Tremaine (1994)]{el044} Elmegreen B. G., Scalo J., 2004,  ARA\&A,  42,  211
\bibitem[Binney \& Tremaine (1994)]{el02} Elmegreen, B. G. 2002, ApJ, 577, 206
\bibitem[Downes \& Solomon (1998)]{el08} Escala A., Larson R. B., 2008,  \apj,  685,  L31
\bibitem[Binney \& Tremaine (1994)]{ga05} Gallagher J. S., Garnavich P. M., Berlind P., Challis P., Jha S., Kirshner 
R. P., 2005, \apj,  634,  210
\bibitem[Binney \& Tremaine (1994)]{gi98} Giraud, E. 1998, AJ, 116, 1125
\bibitem[Binney \& Tremaine (1987)]{az98} Gammie C. F.,  2001, \apj,  553,  174
\bibitem[Binney \& Tremaine (1994)]{gl65} Goldreich P., Lynden-Bell
  D., 1965,  MNRAS, 130,  125
\bibitem[Binney \& Tremaine (1994)]{he03} Helfer, T.T., et al. 2003,
  ApJS, 145, 259
\bibitem[Binney \& Tremaine (1994)]{he08} Hsieh P.-Y., Matsushita S., Lim J., Kohno K., Sawada-Satoh S., 2008, \apj,  683,  70
\bibitem[Binney \& Tremaine (1994)]{jo05} Jogee S., Scoville N.,
  Kenney J. D. P., 2005, \apj,  630,  837
\bibitem[Kennicutt (1998)]{ken89} Kennicutt, R. C. 1989, \apj, 344, 685
\bibitem[Kennicutt (1998)]{ken98} Kennicutt, R. C. 1998, \apj, 498, 541
\bibitem[Binney \& Tremaine (1994)]{ke87} Kent, S.M. 1987, AJ, 93, 816
\bibitem[Binney \& Tremaine (1994)]{ko01} Krumholz M. R., McKee C. F., 2005, \apj,  630,  250

\bibitem[Binney \& Tremaine (1994)]{kk01} Kravtsov, A. V. 2003, ApJ, 590, L1
\bibitem[Binney \& Tremaine (1994)]{ll03} Larsen S. S., 2002, AJ,  124,  1393
\bibitem[Binney \& Tremaine (1994)]{la82} Larson R. B., 1982, \mnras,  200,  159
\bibitem[Binney \& Tremaine (1994)]{jre45} Mac Low, M.-M., Balsara, D. S., Kim, J., \& de Avillez, M. A. 2005, ApJ, 626, 864
\bibitem[Kennicutt (1998)]{ken01} Martin C. L., Kennicutt R. C., Jr.,
  2001,  Astrophysical Journal,  555,  301
\bibitem[Binney \& Tremaine (1994)]{la81} Mauersberger R., Henkel C.,
  Wielebinski R., Wiklind T., Reuter H.-P., 1996, A\&A,  305,  421
\bibitem[Binney \& Tremaine (1994)]{np97} Nordlund, A.,  Padoan, P. 1998, in Interstellar Turbulence, ed. J. Franco \& A. Carraminana (Cambridge: Cambridge Univ. Press), 218
\bibitem[Binney \& Tremaine (1994)]{la81} Padoan, P., Nordlund, A.,
  Johns, B. J. T. 1997, MNRAS, 288, 145
\bibitem[Binney \& Tremaine (1994)]{re4} Padoan, P., \& Nordlund,
  A. 2002, ApJ, 576, 870
\bibitem[Binney \& Tremaine (1994)]{pe4} P\'erez-Torres M. A., Alberdi
  A., 2007, \mnras,  379, 275
\bibitem[Binney \& Tremaine (1994)]{pe4} Pisano D. J., Wilcots E. M., Elmegreen B. G., 1998,  \apj,  115,  975
\bibitem[Binney \& Tremaine (1994)]{la85} Rafikov R. R., 2009,  ArXiv
  e-prints,   arXiv:0901.4739
\bibitem[Binney \& Tremaine (1994)]{sa98} Scalo, J., V\'azquez-Semadeni, E., Chappell, D., \& Passot, T. 1998, ApJ, 504, 835
\bibitem[Binney \& Tremaine (1994)]{sa59} Schmidt, 1959
\bibitem[Binney \& Tremaine (1994)]{ti59} Thilker D. A., et al. 2007,
  ApJ Supplement Series,  173,  572 
\bibitem[Binney \& Tremaine (1994)]{to64} Toomre, A. 1964, ApJ, 139, 1217
\bibitem[Binney \& Tremaine (1994)]{va94} V\'azquez-Semadeni, E. 1994, ApJ, 423, 681
\bibitem[Binney \& Tremaine (1994)]{wn01}  Wada, K., \& Norman,
  C. A. 2001, ApJ, 547, 172
\bibitem[Binney \& Tremaine (1994)]{wn07}  Wada, K., \& Norman,
  C. A. 2007, ApJ, 660, 276
\bibitem[Binney \& Tremaine (1994)]{wn09} Wang, P., \& Abel, T. 2009, ApJ, 696, 96
\bibitem[Binney \& Tremaine (1994)]{van01} Weidner  C., Kroupa  P., Larsen  S. S., 2004, MNRAS, 350, 1503
\end{thebibliography}
\end{document}